# Effect of electric/magnetic field on pinned/biased moments at the interfaces of magnetic superlattices

P. Padhan, and W. Prellier

Laboratoire CRISMAT, CNRS UMR 6508, ENSICAEN,

6 Bd du Marehal Juin, F-14050 Caen Cedex, FRANCE.

## Abstract

We have observed the pinned/biased moments in the superlattices consisting of ferromagnetic (FM) $SrRuO_3$ (SRO) and antiferromagnetic (AFM) $SrMnO_3$ (SMO) bilayer. The alternate stacking of SRO and SMO leading to a low field positive magnetoresistance with enhanced hysteretic field dependent magnetoresistance under the application of the out-of-plane magnetic field. We attribute these effects to the observed biased/pinned magnetic moments in the SRO layer in the vicinity of the interfaces. In addition, the biased/pinned moments can be oriented under the application of either the out-of-plane magnetic field or a combination of out-of-plane magnetic field and in-plane electric field. These results will bring new insights in the understanding of the coupling at the AFM/FM interface which can be useful for creating new exotic phenomena at the interfaces of the multilayer.

# I. INTRODUCTION

The magnetic structure based on 3d - transition metal compounds exhibit overdamped oscillatory exchange coupling[1,2,3], unidirectional magnetic anisotropy such as horizontal/vertical magnetic hysteresis loop shifts[4,5,6,7] and enhanced coercivity[8,9]. In antiferromagnetic(AFM)/ferromagnetic(FM) heterostructures the coercivity and unidirectional anisotropy are enhanced, in contrast to corresponding free magnetic multilayers. The shift in the hysteresis loop i.e. the unidirectional anisotropy of the ferromagnetic layer is attributed to the interfacial effects like exchange coupling. In general, exchange bias is established through field cooling (FC) in the film plane where the magnetic easy axis of soft ferromagnetic materials normally lies in plane due to the shape anisotropy. However, the exchange bias effect was also observed in the zero-field-cooled (ZFC) ferromagnetic bilayer with the application of external magnetic field prior to the measurement of the hysteresis loop[10]. Apart from the industrial applications, exchange bias effect is extensively used to pin the moments in the ferromagnetic layer in the magnetic devices like spin-valve or magnetic tunneling junction. Most of the theoretical models deal with exchange bias, consider that these effects originate from the formation of magnetic domains, either in the AFM or in the FM layer.

In previous articles, the multilayer consisting of FM $SrRuO_3$ (SRO) and AFM $SrMnO_3$ (SMO) were grown to study the exchange coupling[6,7] at the interfaces of SRO and SMO. Below a certain critical magnetic field, we observed that the out-of-plane field cooled hysteresis loop is shifted along the magnetization axis. In this article, we compare the transport and magnetic measurements of SRO/SMO superlattices in order to understand the physical process due to the out-of-plane magnetic and in-plane biased electric field. The results show that, either the magnetic field and/or the combination of magnetic and electric field has a strong influence on the exchange coupling at the SRO-SMO interfaces, in addition to the cooling field. Moreover, these external perturbations align the pin spin in the FM regions and lead to a strong magnetic anisotropy.

## II. EXPERIMENT

The SRO/SMO superlattices were grown on (001)-oriented $SrTiO_3$ (STO) substrates by pulsed laser deposition using a KrF excimer laser. The details of optimized deposition conditions are described elsewhere[11]. Briefly, the superlattice structures were synthesized by repeating 15 times, the bilayer comprising of 20-(unit cell, u.c.) SRO and n-(u.c.) SMO, with n taking integer values from 1 to 20. In all samples, the bottom layer is SRO and the modulation structure was covered with 20 u.c. SRO to keep the structure of the topmost SMO layer stable. The superlattices were characterized by magnetotransport and magnetic measurements, in addition to x-ray diffraction. The magnetization measurements were carried out by cooling the sample below room temperature in the presence or absence of magnetic fields along the [001] direction of the STO substrate. The orientation of the magnetic field during the field-cooled measurements remains similar to that of the cooling field. Note that SRO is a metallic ferromagnet, with a Curie temperature ($T_C$) ~ 160 K in its bulk form[12] and SMO is an AFM with a Neel temperature ($T_N$) close to 260 K [13].

## III. RESULTS

Fig. 1 shows the low angle $\theta$ - $2\theta$ x-ray scan of the superlattices with n = 20 deposited on (001) STO. It also includes the simulated spectrum obtained from the quantitative refinement of the superlattice structure using DIFFaX[14] program. The simulated profile is in good agreement with the measured $\theta$ - $2\theta$ x-ray scan with respect to the position of Bragg's peaks and their relative diffracted intensity ratio. The low angle peak position ($\theta$) of the modulated structure is related to the period ($\Lambda$) of chemical modulation and its averaged index of refraction. Miceli et. al. [15] have indexed the low angle peak positions with the modulation period as:

1. $\sin^2\theta = [m\Lambda/2\Lambda]^2 + 2\ \overline{\delta}_S$

Where m, Λ and (1 - $\overline{\delta}_S$) are the order of reflection, the x - ray wave length and the real part of the average index of refraction of the superlattice, respectively. The low angle peak positions of this sample follow the eq. 1 (see inset of Fig. 1) indicating a good quality of the superlattice. The value of $\overline{\delta}_S$ obtained from the intercept of $m^2$ vs. $\sin^2\theta$ plot is 2.8 x $10^{-5}$ which is close to the value of $\overline{\delta}_S$ that gives a significant deviation from the Bragg's law at 2θ ≤ 3°[16]. The value of superlattice period has been calculated from the slope and it agrees well with the value obtained at higher angle diffraction profile[11].

These samples show temperature dependent zero-field resistivity similar to that of the thin film of SRO[11], also their zero-field resistivity drops in the presence of 7 tesla magnetic field. At 10 K the magnetoresistance (MR) under a 7 tesla out-of-plane magnetic field for the sample with n = 5 and 20 is -1.27 % and -1.15 % respectively. The field-dependent magnetoresistance (MR-H) of these two superlattices measured at 10 K with the magnetic field perpendicular to the film plane is shown in the Fig. 2. The hysteretic nature of field dependent magnetoresistance (MR-H) at 10 K of these samples is enhanced, though the MR is suppressed compared to single layer SRO film on STO, with the out-of-plane magnetic field. In addition, some other interesting features are observed in the (MR-H) loops of these samples. First, the MR is negligibly small and constant under a field below 1 tesla, at the beginning of the field increasing branch. Second, the MR drops sharply at a switching field for the sample with n = 5, but a second order-like-transition appears for the sample with n = 20. Third, the low field MR is positive. Fourth, the high field MR is negative and increases monotonically with magnetic field. Finally the (MR-H) loops are symmetric to the MR axis. This suggests that a magnetic field lower than 1 tesla is not sufficient to alter the spin dependent scattering of the sample. This could be due to the shorting of top SRO layer. However, the variation of MR at the switching field corresponding to the coercive field is related to the variation of the exchange coupling with the AFM SMO layer thickness. The hysteretic effect is attributed to the spin dependent switching of charge carriers with the localized moments in SRO below the switching field. The hysteretic effect in these samples might be due to the magnetic field or biased electric field or both.

The resistance of the sample in the field increasing branch for a magnetic field below the switching field is higher compare to the field decreasing branch which suggests that the irreversible modification of the orientation of the localized magnetic moments. Similarly some irreversible changes observed in the field dependent magnetization is shown in the Fig. 3. To picture the variation of the magnetic moments with the external magnetic field in the superlattices, we have measured the magnetization of the superlattice with 5 u.c. thick SMO layer. The magnetization was measured for several successive cycles, where the negative magnetic field in the range was fixed at 1 tesla and the range of the positive magnetic field was varied from 1 to 3 tesla with a step of 1 tesla. The negligibly small change in the magnetization of the field increasing and field decreasing paths suggests the manifestation of the unidirectional pinning in the initial randomly pinned state of the sample which is partially responsible for the hysteretic nature of the (MR-H) loop below the switching field.

To understand the effect of the magnetic and/or electric field on the localized magnetic moments, the minor magnetic hysteresis loops were measured below the switching field in the presence of a magnetic field, electric field and their combination. After cooling the sample down to 10 K, in the absence of magnetic and electric fields, the minor hysteresis loops were measured between ± 1 tesla magnetic field oriented along the [001] direction of the substrate. The minor hysteresis loops were measured at four different conditions to study the effects of magnetic and electric fields. The different conditions are: at 10 K the sample was subjected to zero, - 4 tesla magnetic field, and - 4 tesla magnetic field with ± 1.2 Volt/cm electric field. These fields were set to zero before the measurement of the hysteresis loop. The minor ZFC hysteresis loop shown in the Fig. 4(a) is symmetric around both the field as well as magnetization axis. However, as the ZFC sample was subjected to - 4 tesla magnetic field (a field above the switching field) before the field sweep, the hysteresis loop shifts along the negative magnetization axis. This shift of the magnetic hysteresis loop is enhanced as the sample was subjected to the combination of magnetic as well as electric field (Fig. 4a). This suggests that the moment can be biased by the application of either a magnetic field or a magnetic field with an electric field. However, the shift in the ZFC hysteresis loop along the magnetization axis is negligibly small when the sample was subjected to only an electric field (Fig. 4b).

## IV. DISCUSSION

Since the conductivity of SRO is larger compare to the SMO and top layer is SRO, the in-plane transport in these superlattices is dominated by SRO. However, its physical properties also depend on the other degree of freedoms those comprises of magnetic interaction at the interfaces and the effective strain of the structure. The lower value of anomaly in the temperature and the low-temperature insulator like behavior of the temperature dependent resistivity compare to that of the bulk SRO suggest the presence of interfacial magnetic modification[6,7,17,18,19]. This magnetic modification can be viewed as the canted state of the moment near the interface in SRO layer. This effect is confirmed as the saturation moment in the (M-H) loop of the sample is much lower than the saturation moment of SRO. However, the insulator like behavior at the lowest temperature is the signature of a strong localized moment whose orientation is not altered even in 7 tesla magnetic field. These localized moments are responsible for the increase of the spin dependent scattering at the lowest temperature. Moreover, the features of the magnetic transports are not significant in the temperature-dependent resistivity.

To explain this biasing mechanism, we have used the model given by Meiklejohn[20]. In this model, the energy per unit interface area, assuming the coherent rotation of magnetization, can be expressed as:

2. $E = - H M_{SRO} t_{SRO} \cos(\theta-\beta) + K_{SRO} t_{SRO} \sin^2(\beta) + K_{SMO} t_{SMO} \sin^2(\alpha) - J_{SRO\text{-}SMO} \cos(\beta-\alpha)$,

where H is the applied field, $M_{SRO}$, $t_{SRO}$ and $K_{SRO}$ are saturation magnetization, thickness and anisotropy of the SRO layer, respectively. $t_{SMO}$ is the thickness and $K_{SMO}$ is anisotropy of the SMO layer and $J_{SRO\text{-}SMO}$ is the exchange coupling constant at the interface of SRO and SMO. $\theta$ is the angle between the easy axis of SRO and the direction of the magnetic field, $\beta$ is the angle between the direction of the magnetization and the easy axis of the SRO layer and $\alpha$ is the angle between the easy axis of SMO and its AFM sublattice magnetization.

In the ZFC magnetic hysteresis loop with a magnetic field range ± 5 tesla, the in-plane magnetization of the superlattice gradually increases as the magnetic field increases and becomes larger than the calculated value (1.6 $\mu_B$/Ru), based on the only contribution from SRO layer while the out-of-plane hysteresis loop shows a clear saturation magnetization and saturation field[6]. The major contribution of this anisotropy behavior is from the strong anisotropy of the SMO layers and the additional periodicity of the magnetic layer along the out-of-plane direction of the sample. Thus, the easy axis of SMO is along the [100] direction of STO and the easy axis of SRO is along the [001] direction of STO with negligibly small magnetic anisotropy ($K_{SRO}$)[7]. Though the easy axis of SRO in SRO-SMO superlattice is along the [001] direction of the substrate, the exchange coupling between SRO-SMO favors the in-plane orientation of the SRO moment for the minor hysteresis loop, with a magnetic field range ± 1 tesla of these superlattices[6,7]. The possible angular configuration of the magnetic field and the magnetization of the SRO and the sublattices of SMO in the superlattice expected from the magnetic properties of these samples is shown in the Fig. 5. The ideal superlattice structure consist of FM and AFM layers, however, in our earlier study we have modeled the ideal structure as a repetition of AFM/(pin)/FM(free)/(pin) units[6]. In the absence of the magnetic field, the last two terms in eq. 2 contribute to the total coupling energy (E). The application of an external magnetic field induced a competing effect between the Zeeman energy term with the anisotropy energy of SMO and the exchange energy term in eq. 2. Since the Zeeman energy varies with the magnitude of the external magnetic field, the application of the magnetic field will also vary the exchange energy at the interfaces. Thus, the values of $\theta$ and $\beta$ shown in the Fig. 5 will change with the variation in the magnitude of the out-of-plane magnetic field. When the sample is subjected to the magnetic field larger than the critical pinning energy the Zeeman energy dominant exchange energy leading to an oriented pin layer. So, in this magnetic system the coupling energy between the SRO and SMO is strongly depends on the Zeeman energy. Consequently, the effect of the magnetic field upon the unidirectional anisotropy of the magnetization is significant. The presence of electric field and magnetic field produce spin polarized carriers in SRO[21]. Since the electric and magnetic field are perpendicular to each other and the spacing between the electrodes of the electric fields is several orders of the mean free path, the spin current carriers are expected to pass through the interfaces. This may affect the physical processes like spin-diffusion, spin-accumulation and spin-dependent reflection due to the bias/pin moment at the interfaces. Also in the magnetically inhomogeneous interfaces, the spin current carries an exchange field that contributes to the local moments of a layer and forces its magnetization to align along the spin polarization of the conduction electrons[22].

## V. CONCLUSIONS

In conclusion, we have observed pin/bias moment in the FM/AFM superlattices consisting of ferromagnetic SRO and antiferromagnetic SMO bilayer. The strong exchange coupling at the SRO-SMO interfaces pinned/biased magnetic moments randomly at the SRO layer in the vicinity of the interfaces. This random pinning/biasing can be made to unidirectional anisotropy by the application of magnetic field or combination of magnetic and electric field, below the switching magnetic field. This manifestation of unidirectional anisotropy moment by the application of magnetic field is a consequence of the enhanced Zeeman energy or the spin-accumulation at the interfaces. The aligned pin/bias moments and the increase in pinning strength are realized in the field dependent resistivity measurements. Also, the flipping of pinned/biased moments is reflected in the field dependent magnetoresistance with a sharp change in MR (positive MR changes to negative MR). Since correlated electron physics is now seeking for new functions created at interfaces, which may induce new exotic properties and phenomena, these results may provide some ideas about the interface issue of magnetic oxides in general and perovskite oxides in the present case.

## ACKNOWLEDGMENTS:

We greatly acknowledge the financial support of the Centre Franco-Indien pour la Promotion de la Recherche Avancee/Indo-French Centre for the Promotion of Advance Research (CEFIPRA/IFCPAR) under Project N°2808-1. We also thank Dr. H. Eng for careful reading of the article.

Figure captions:

Fig. 1: Measured and simulated low angle θ-2θ x-ray diffraction spectra for the superlattice with n = 20. Inset shows the $m^2$ vs. $\sin^2\theta$ plot of the samples with n = 5 and 20.

Fig. 2: Zero-field-cooled magnetoresistance (MR= (R(0)-R(H))/R(0)) at various magnetic field oriented along the [001] direction of STO of the superlattice with (a) n = 5 and (b) n= 20. The arrows indicate the directions of the field sweep. The thicker arrow indicates the direction of the field at the beginning.

Fig. 3: Three successive cycles of isothermal (10 K) ZFC hysteresis loop of the superlattice with n = 20 with magnetic field orientation along the [001] direction of the substrate. During the field sweep the negative field was fixed while the positive field was increased in the successive field sweep cycle. The arrows indicate the directions of the field sweep. The thicker arrow indicates the direction of the field at the beginning.

Fig. 4(a): Isothermal (10 K) ZFC magnetization of the superlattice with n = 20 at various fields oriented along the [001] directions of the substrate measured at three different conditions. The different conditions are: at 10 K the ZFC sample was subjected to (i) zero, (ii) - 4 tesla magnetic field and (iii) - 4 tesla magnetic field with ± 1.2 Volt/cm electric field.

Fig. 4(b): Isothermal (10 K) ZFC magnetization of the same superlattice at various fields with same orientation, the sample was subjected to 1.2 Volt/cm electric field prior to the field sweep.

Fig. 5: Schematic representation of the orientation of magnetic field, magnetic moment and magnetic anisotropy of SRO and SMO in the SRO-SMO superlattices.

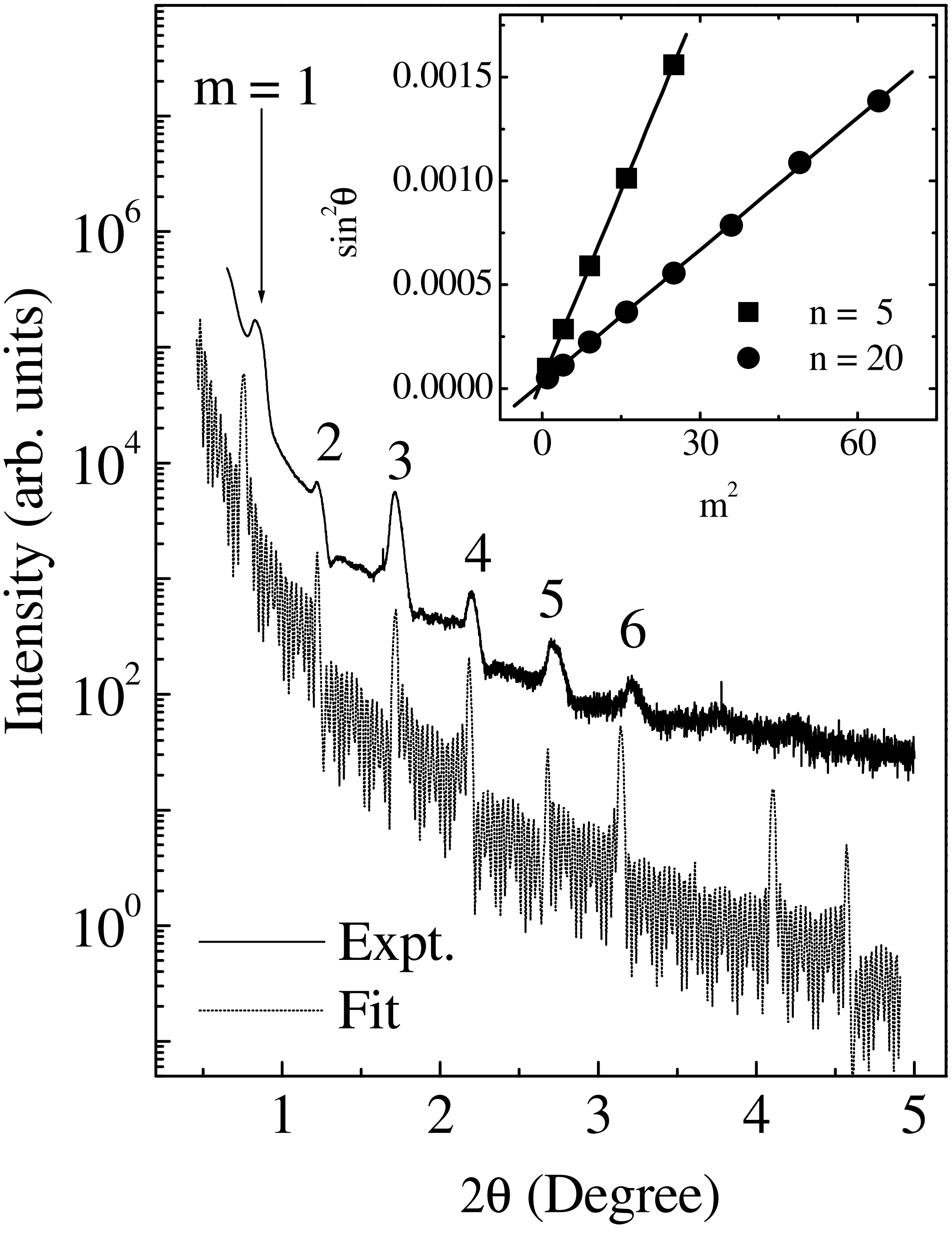

Fig. 1
Padhan and Prellier

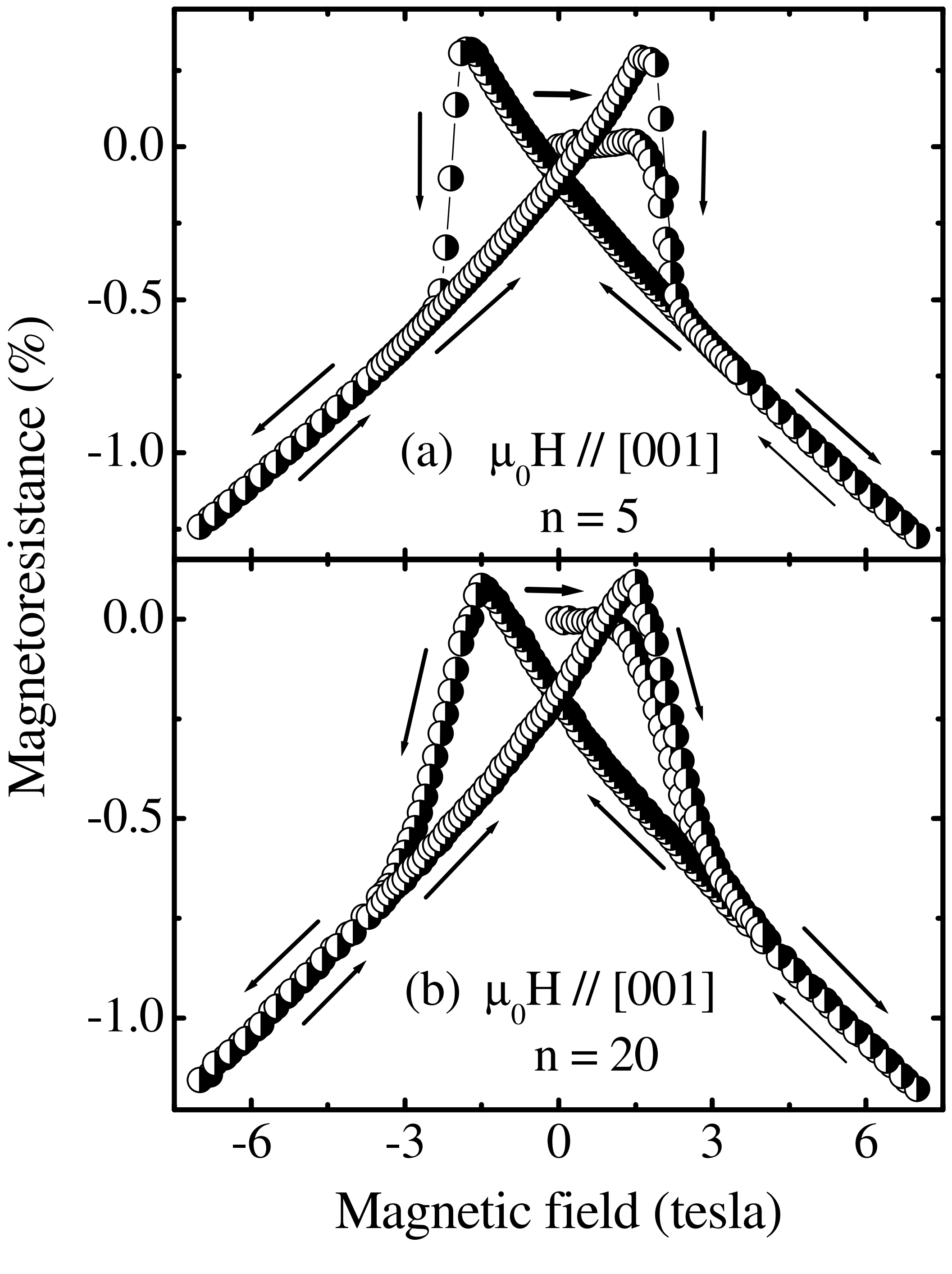

Fig. 2
Padhan and Prellier

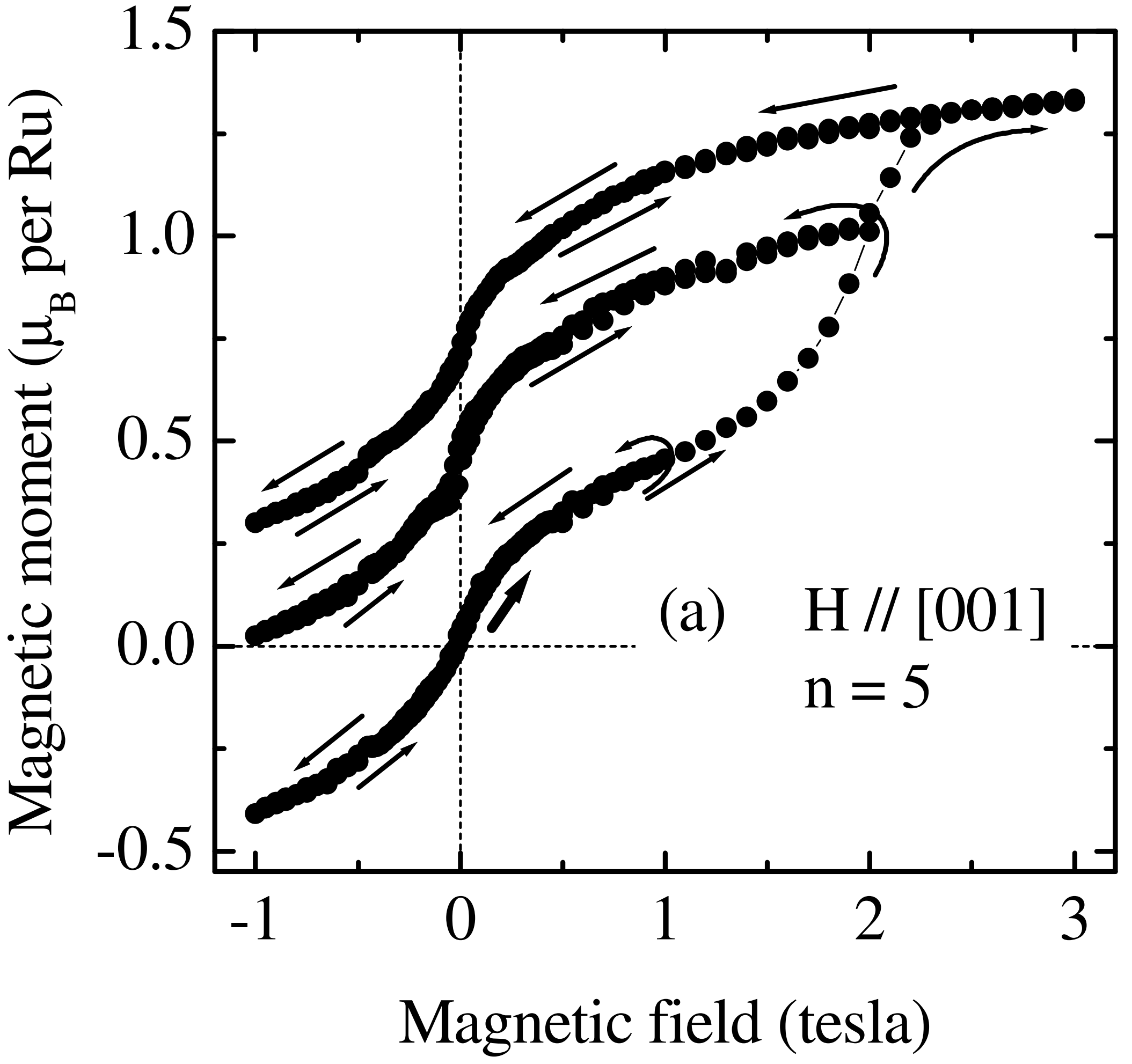

Fig. 3
Padhan and Prellier

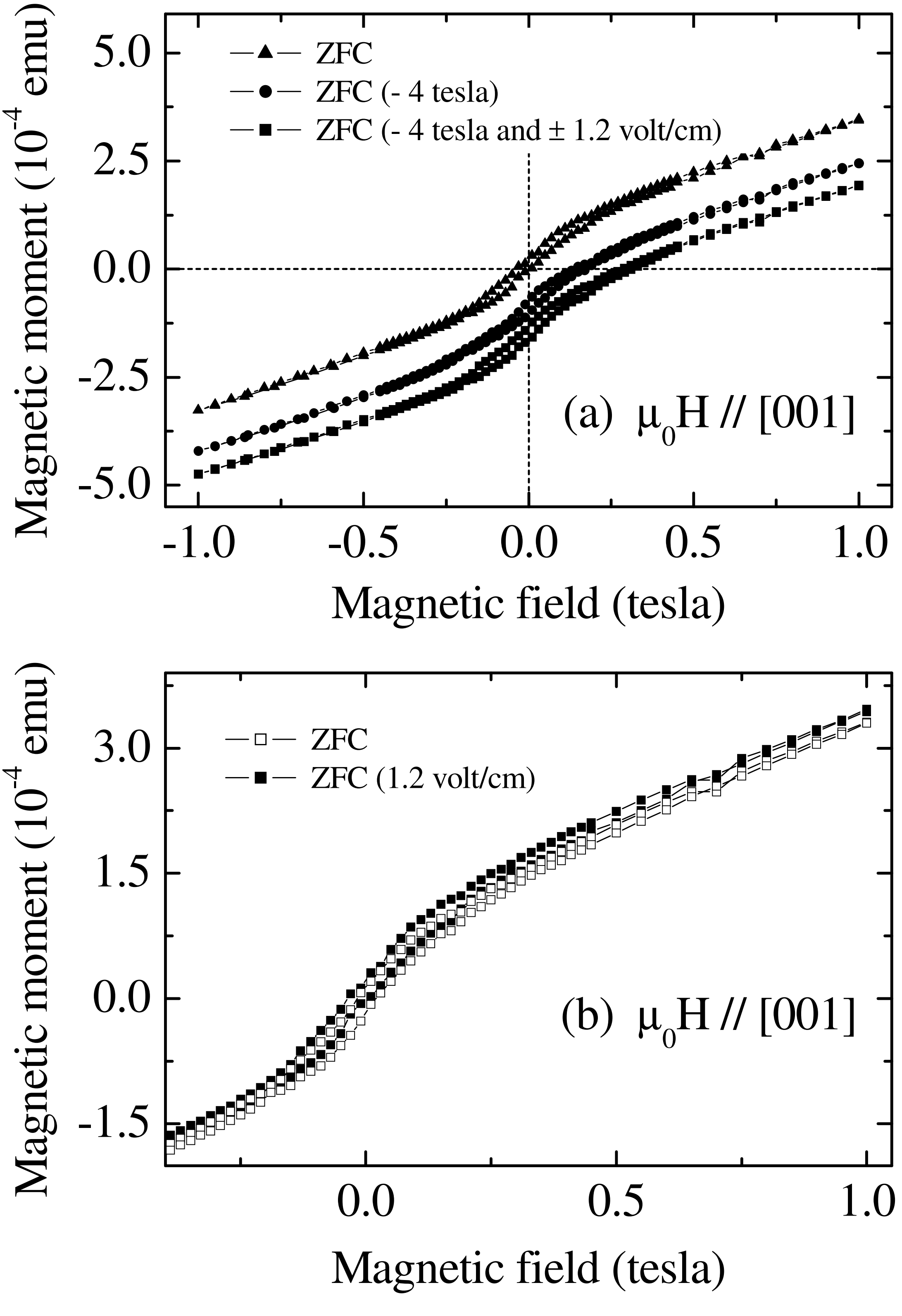

Fig. 4
Padhan and Prellier

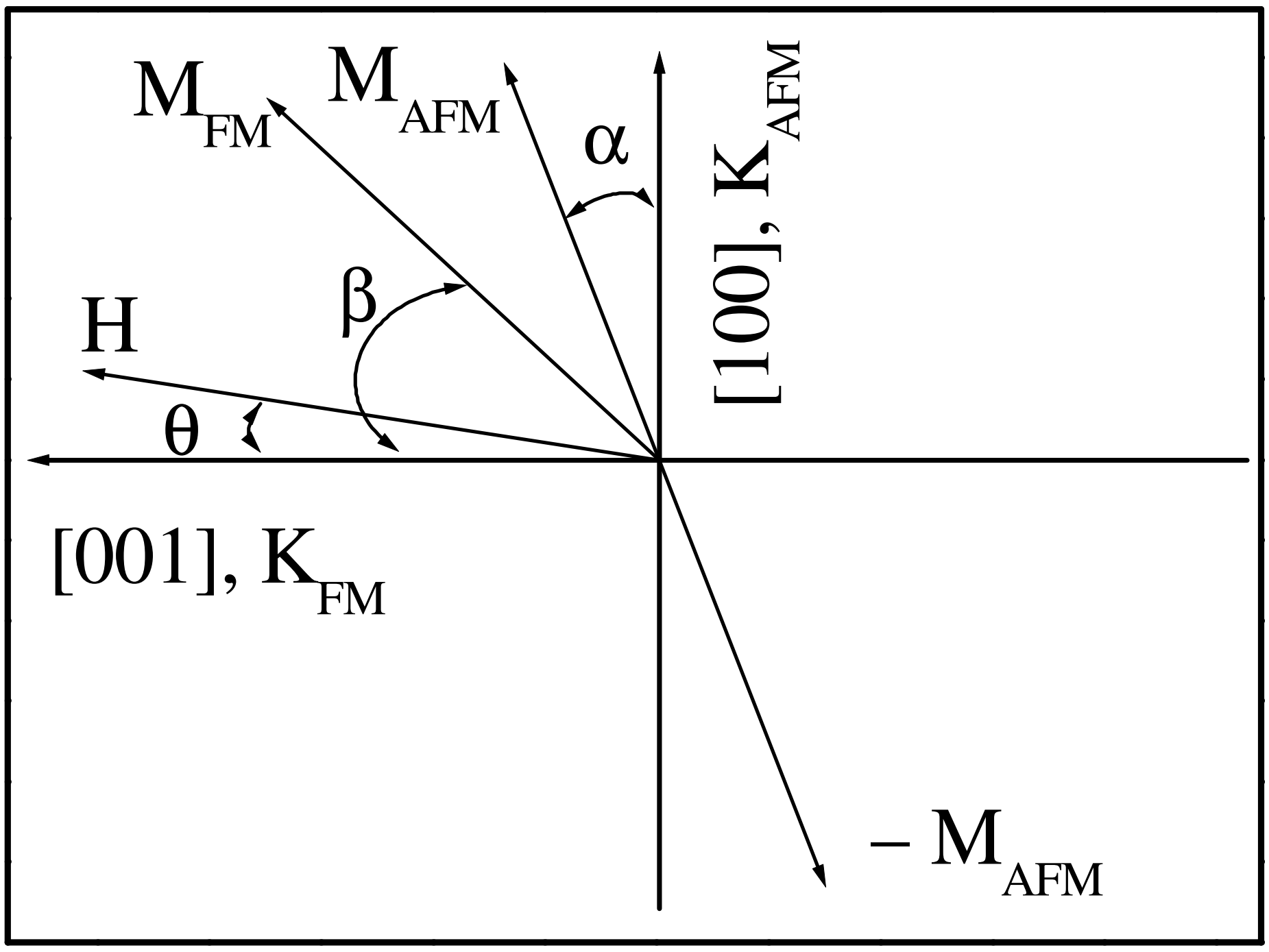

Fig. 5
Padhan and Prellier